\documentclass[superscriptaddress,pre]{revtex4}
\usepackage{graphicx,amsfonts}

\begin{document}

\title{Algebraic coarsening in voter models with intermediate states}

\author{Luca Dall'Asta}
\email{dallasta@ictp.it}
\affiliation{Abdus Salam International Center for Theoretical Physics, Strada
  Costiera 11, 34014 Trieste, Italy}
\author{Tobias Galla}
\email{tobias.galla@manchester.ac.uk}
\affiliation{Theoretical Physics Group, School of Physics and Astronomy, University of Manchester, Manchester M139PL, United Kingdom}
\affiliation{Abdus Salam International Center for Theoretical Physics, Strada
  Costiera 11, 34014 Trieste, Italy}


\begin{abstract}
The introduction of intermediate states in the dynamics of the voter
model modifies the ordering process and restores an effective surface
tension. The logarithmic coarsening of the conventional voter model in two
dimensions is eliminated in favour of an algebraic decay of the
density of interfaces with time, compatible with Model A dynamics at
low temperatures. This phenomenon is addressed by deriving
Langevin equations for the dynamics of appropriately defined
continuous fields. These equations are analyzed using field
theoretical arguments and by means of a recently proposed numerical
technique for the integration of stochastic equations with
multiplicative noise. We find good agreement with lattice simulations
of the microscopic model.
\end{abstract}


\maketitle

\section{Introduction}
The theory of phase ordering dynamics is a prominent tool to study and
classify the non-equilibrium behaviour of interacting particle systems
\cite{B94}. Applications can be found in very different fields,
ranging from the study of binary mixtures \cite{GSS83} over
reaction-diffusion systems \cite{H00} to models of social dynamics and
opinion spreading \cite{CFL07}.

In many physical systems, phase ordering occurs through defect
annihilation and domain coarsening, as it can be observed for example
in the Ising model after a quench from the disordered phase into the
ordered one at low temperatures \cite{B94}. Here, the typical length
scale $L(t)$ of clusters of aligned spins grows algebraically with
time, following the general behaviour $L(t)\sim t^{1/z}$, where
e.g. $z=2$ for non-conserved dynamics.  The numerical value of the
{\em dynamical exponent} $z$ for any given model here depends on universal
features of the model, such as symmetries and conservation laws, and
on the relevant external parameters.\\ The low-temperature ordering in
Ising-like models exhibits another very general property: the
existence of surface tension between domains, favouring the formation
of smooth interfaces minimizing the curvature.  This motivates the
notion of {\em curvature-driven} coarsening.  In presence of smooth
interfaces, we also expect the density of interfaces $n(t)$ to be
inversely proportional to the domain length $n(t) \sim {L}^{-1}$, thus
decaying with similar algebraic law $n(t) \sim t^{-\delta}$ with
$\delta = 1/z$.

Other systems exist in which coarsening takes place without surface
tension \cite{DCCH01}; the growth of clusters is there purely
diffusive, driven by interfacial noise. A typical example is provided
by the voter model \cite{L85}, in which spins can take one of two
possible states, and flip with a probability linearly proportional to
the fraction of unlike neighbors. In dimension $d=1$, the voter
dynamics is the same as zero-temperature Glauber dynamics of the Ising
model \cite{G63}, but increasing space dimensionality the ordering
slows down and becomes logarithmic in $d=2$ ($n(t) \sim 1/\log(t)$),
whereas an infinite system does not order for $d>d_{c} =2$.  The
notion of logarithmic coarsening in two dimensions here refers to the
behaviour of the density of interfaces with time, whereas the
dynamical exponent, relating to the correlation length and properties
of the dynamical structure factor, is still found to be $z=2$
\cite{SS88}. This is due to the critical nature of the voter model, as
unveiled using field theoretic approaches \cite{P86,L94}. The voter
dynamics corresponds to a massless field theory without field
renormalization (from which $z=2$), but it admits renormalization of
other quantities, including the density of interfaces $n(t)$.  Exact
renormalization group calculations predict a behavior that agrees
perfectly with both analytical solutions and numerical simulations
\cite{K92,FK96}.

Recent research in opinion dynamics and sociophysics has amplified the
interest in these two distinct classes of non-equilibrium systems, and
has prompted their use in the study of the collective behavior of
social individuals \cite{CFL07}.  For instance, voter-like models with
multiple states have been used to model groups of political parties
\cite{VKR03,DSIW06}, the spreading of linguistic conventions
\cite{CES06,BDBL06,BDBL07}, while the introduction of counters
\cite{DC07} or inertia in the spin dynamics \cite{STS07} have been
used to describe the effect of learning and memory in social
interactions.  Interestingly, the mere introduction of an intermediate
state, without any major modification of the voter dynamics itself,
has been seen to be sufficient to avoid logarithmic coarsening in two
dimensions, restoring an algebraic decay of the density of interfaces and
an effective surface tension \cite{DC07,CES06}.  As we will show
below, the qualitative asymptotic behavior does not change if one
allows for more than one intermediate state, provided that only two
absorbing states are present and that they correspond to the extreme
states.

A better understanding of this intriguing phenomenological picture can
be obtained by regarding the voter model as a special point of a
larger class of non-equilibrium systems with two absorbing states,
exhibiting a $\mathbb{Z}_{2}$-symmetry, see \cite{OMS93,
DG99,DCCH01}. Starting from this viewpoint, Al Hammal et
al. \cite{ACDM05} have derived a simple phenomenological field theory,
expressed in terms of a non-linear Langevin equation, which captures
the essential characteristics of the dynamical behavior of these
models, and which faithfully reproduces the logarithmic coarsening
dynamics of the voter model.

The aim of our work is to provide a theoretical explanation for the
restoration of curvature-driven coarsening by intermediate states, and
to corroborate the microscopic view with a continuum approach able to
clarify the link with known results on models in the the so-called
{\em generalized voter} (GV) class \cite{DCCH01,ACDM05}.  In the
following we show analytically that voter models with intermediate
states (this includes e.g. Naming Games \cite{CES06,BDBL06}) can be
regarded as members of this class, indirectly providing an example of
non-trivial microscopic models belonging to the GV class. We will
derive phenomenological equations for the $3$-state voter model, and
describe how such equations are, within certain limits, useful to describe the dynamics of
models with a larger number of intermediate states as well.  These
Langevin equations are studied both numerically and more formally
using field theoretical arguments. We show that despite the presence
of interfacial noise, in two dimensions the asymptotic behavior is
compatible with the low (but non-zero) temperature regime of Model A
relaxational dynamics.  We also show that, unlike in the Ising model,
interface roughening is not expected to occur in higher dimensions.

\section{Voter models with and without intermediate states}
In the voter model \cite{L85} spins are arranged on a network with one
spin at each node, and each spin can assume two possible values,
e.g. $\pm1$. At each step of the microscopic dynamics one spin is
selected at random, and subsequently assumes the state of one randomly
chosen nearest neighbour. We introduce an intermediate $0$-state,
maintaining a dynamics similar to that of the voter model. One spin
$s$ and one of its neighbours $s'$ are chosen randomly at each time
step. If the spin selected is in state $s=+1$ and the neighbouring one
in state $s'=-1$, then $s$ assumes the value $s=0$. Similarly for
$s=-1,s'=1$ the first spin will be set to $s=0$. If $s=1$ and $s'=0$,
then $s$ will assume the state $s=0$ with probability $1/2$ and remain
at $s=1$ otherwise. Similarly for $s=-1,s'=0$. If $s=0$, then $s$ will
adopt the state $s'$ if $s'=\pm 1$, and will flip to either $s=1$ or
$s=-1$ with equal probability if $s'=0$. The exact dynamical iteration
prescriptions and transition rates for the voter model and the
modified $3$-state model are summarised in Table~\ref{table1}.  Note
that with these rules intermediate states do not persist in time, and
the only possible absorbing states are those with either all spins up
or all spins down. In this work we will focus on models defined on
$d$-dimensional regular lattices of lateral size $L$, and most of the
numerical analysis is carried out in the two-dimensional
case, where we impose periodic boundary conditions. Interaction occurs with the set of the four nearest neighbours
of any given spin.

\begin{table}
\centering{
\begin{tabular}{|c|c|}
\hline
Model & Transition rates \\
\hline
$2$-state model  &  $p_{\pm \to \mp} = f_{\mp}$  \\
\hline
$3$-state model  &  $p_{\pm \to 0} = f_{\mp} + \frac{ f_{0}}{2}$, ~~~~   $p_{0 \to \pm} = f_{\pm} + \frac{ f_{0}}{2}$\\
\hline
\end{tabular}
}
\caption{Transition rates of the two models considered. The $2$-state (spin $1/2$) model is the usual voter model, while the $3$-state model (spin $1$) is the one with an intermediate state. The quantities $f_{\pm},f_{0}$ are the fractions of $\pm,0$ spins among the $4$ nearest neighbours of a given spin.}
\label{table1}
\end{table} 

\begin{figure}
\centerline{
\includegraphics*[width=0.5\textwidth]{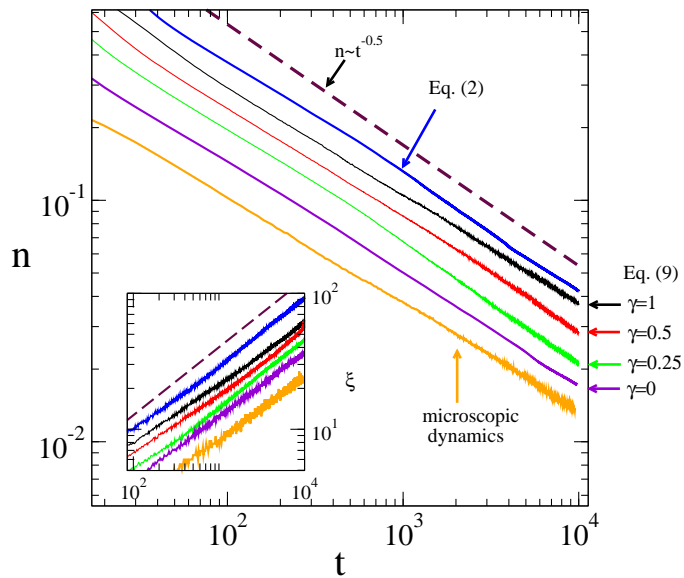}}
\caption{(Colour on-line) Density of intermediate states (main panel) for the $3$-state model. Results from one sample of a direct microscopic simulation of a system of size $2000^2$ are shown, along with measurements from numerical integration of Eqs. (\ref{3voter-eq}), where we have ignored the noise terms (see text; system size is $500^2$, averages over $10$ runs with independent random initial conditions), and from Eq. (\ref{eq:eqphi}) at $a=1/2,D=1$ and different noise strengths $\gamma$ (curves at $\gamma>0$ are obtained from systems of size $500^2$, integration method as explained in the main text, averages over $10$ samples are taken; results for $\gamma=0$ are from Euler-forward integration, time-stepping of $0.05$, $5$ samples of size $1000^2$). The inset shows the behaviour of the correlation length $\xi$ of the field $\phi$ (obtained from the circularly averaged correlation function). All curves have been rescaled by arbitrary constant factors for optical convenience. One unit of time corresponds to one update per spin on average.}\label{fig:micro}
\end{figure}

\begin{figure}
\centerline{
\includegraphics*[width=0.4\textwidth]{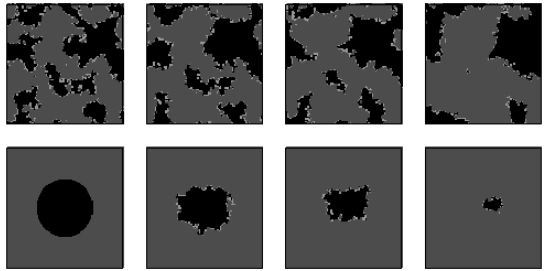}}
\caption{ Snapshots of the time evolution of the model with intermediate states. Sites $i$ with $s_i=-1,0,1$ correspond to black, white and grey respectively. Data is from simulations of the microscopic model, only a region of $N=100^2$ sites is shown. The run in the upper row started from random initial conditions $s_i=\pm 1$, snapshots are taken after $100, 200, 400$ and $800$ sweeps over the system. Lower row shows evolution of a droplet (snapshots are taken at time $t=0$ and then after $500,1000$ and $2000$ sweeps).}\label{fig:droplet}
\end{figure}

\begin{figure}
\centerline{
\includegraphics*[width=0.3\textwidth]{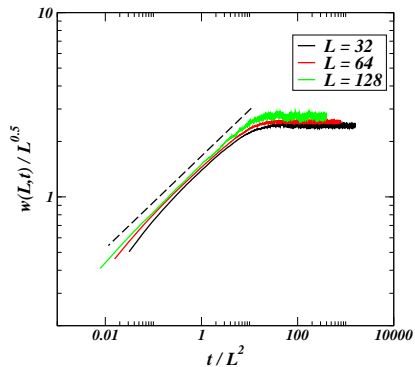}}
\caption{(Colour on-line) Data collapse for the interface width $w(L,t)$. The time is
rescaled by $L^{z}$ (having assumed $z=2$) and the width is rescaled
by $L^{\alpha}$ with $\alpha = 0.5$. This generates an initial
power law growth $t^{\beta}$ with $\beta \approx 0.25$ (dashed
line). This rescaling is thus consistent with the Edwards-Wilkinson
universality class of surface growth.}\label{fig:roughness}
\end{figure}

The $2$-state and $3$-state models behave in the same way in one
dimension, since the presence of intermediate states only has the
effect of renormalizing the width of the domain walls by a finite
factor (see \cite{BDBL06,DC07} for similar considerations).  In two
dimensions the behaviour of the model with intermediate states is
instead very different from that of the standard voter model. This can
be seen in Fig. \ref{fig:micro}, where we report the temporal
evolution of the density of intermediate states for the $3$-state
model. We note here that spins in the intermediate state are found at
the interfaces of domains of up and down spins respectively, so that
the density of intermediate states is a measure of the density of
interfaces (or domain boundaries).  In the standard two-dimensional
voter model, the interface density is known to decrease
logarithmically, while the modified ($3$-state) model shows algebraic
decay, i.e. $n(t) \sim t^{-\delta}$, where $\delta$ has been reported
in the range of $0.45$ to $0.5$ \cite{CES06}. The correlation length
$\xi$, which we have computed from the two-point correlation function,
grows with a similar algebraic law (see inset of
Fig. \ref{fig:micro}). Both measures seem to suggest a dynamical
exponent $z = 2$, even if from these numerical values the contribution
of logarithmic corrections cannot be excluded. These were also the
conclusion of previous studies \cite{DC07,CES06}, and are corroborated
by the marginality of voter dynamics in two dimensions.  The point
will be clarified in the next sections by means of a more theoretical
analysis.\\ Fig. \ref{fig:droplet} displays snapshots of the
coarsening dynamics started from a disordered configuration (top), and
during a `droplet experiment' \cite{DCCH01}, in which a circular
domain of up spins is left to evolve in a sea of down spins (bottom).
In the pictures it is visible that despite the voter-like update rule,
the $3$-state model develops an effective surface tension, as already
noted for similar dynamics with intermediate states, such as the
Naming Game model \cite{BDBL06}, the noise-reduced voter model
\cite{DC07}, and models of bilingualism \cite{WM05,CES06}. \\ To
investigate the role of the interfacial noise clearly visible in the
snapshots in Fig. \ref{fig:droplet}, we have analyzed the roughness
properties of the interface between two domains. Let us consider a
flat interface between $+1$ and $-1$ spins and study its evolution in
time under the dynamics of the model (see \cite{BS95} for details of
such experiments in the context of other models). Fig.
\ref{fig:roughness} displays the behaviour of the average fluctuations
of the width of the resulting interface . More precisely, if we call
$h_{i}$ the displacement in the direction perpendicular to the
interface at position $i$ (the latter is measured along the
interface), the width is defined as $W = {\left[
\frac{1}{L}\sum_{i=1}^{L} h_{i}^2 - {\left(\frac{1}{L} \sum_{i=1}^{L}
h_{i} \right)}^{2} \right]}^{1/2}$, where $L$ is the linear size of
the interface.  As the dynamics progresses $W$ assumes a stationary
value asymptotically. Considering systems of different sizes, we find
that this saturation value $W$ does not remain {\em finite} in the
infinite systems, but that instead the plateau value depends on the
linear size of the system $L$. This indicates the presence of
interface roughening phenomena. As in usual surface growth
\cite{BS95}, the time evolution of the surface width exhibits two
stages separated by a `crossover' time $t_{\times}$. The initial
growth follows a power law, $W(L,t) \sim t^{\beta}$, with growth
exponent $\beta$; then for $t \gg t_{\times}$ the width saturates at
$W_{sat}(L) \sim L^{\alpha}$ where $\alpha$ is the {\em roughness
exponent}. Note that in the scaling theory of surface growth these
exponents are related to the dynamic one by $z = \alpha/\beta$. Figure
\ref{fig:roughness} verifies the Family-Vicsek scaling relation
$W(L,t) \sim L^{\alpha} f \left(t / L^{z} \right)$, with $\alpha =
0.5$ and $z = 2$. Accordingly we find $f(u) \sim u^{\beta}$ with
$\beta \approx 0.25$ at small times. The good data collapse generated
by rescaling, suggests that the interface dynamics in our model is
compatible with the Edwards-Wilkinson (EW) universality class of
surface growth \cite{BS95}.  \\In summary, the numerical results
indicate that the introduction of an intermediate state makes the
voter dynamics similar to the low-temperature ($T>0$) scaling of model
A \cite{B90}, both as far as the domain growth and interface
roughening are concerned.

Further generalisation to multiple intermediate states has already
been considered. For instance, in the Naming Game \cite{BDBL06} a
large number of different intermediate states can be created. These
are however all equivalent with no ordering or hierarchy among them,
therefore the effective behavior of the system is expected to be the
same as for $3$-state model.  More interestingly we can define a
voter model with non-equivalent intermediate states: let us consider
$S\geq 2$ states in total, two of them being extremal absorbing
states, and the remaining $S-2$ ones being intermediate states. To
implement this generalisation let us label the states by
$k=1,\dots,S$, with $k=1$ and $k=S$ being the extremal ones. As before
the model is considered on a two-dimensional lattice, and at each
time-step one spin and one of its four nearest neighbours are chosen
at random. Let $s\in\{1,\dots,S\}$ be the state of the chosen spin,
and $s'$ that of the neighbour. Then set $p(s')=(s'-1)/(S-1)$. As a
consequence of the interaction, spin in state $s$ is then set into
state $s+1$ with probability $p(s')\Theta(S-s)$, and into state $s-1$
with probability $(1-p(s'))\Theta(s-1)$. $\Theta(x)$ is here the step
function with $\Theta(x)=1$ for $x>0$ and $\Theta(x)=0$ for $x\leq
0$. It ensures that spins are never modified to take values outside
the set $\{1,\dots,S\}$. As a consequence of the definition of
$p(s')$, linearly increasing from $0$ to $1$ as $s'$ increases from
$1$ to $S$, the extremal states $1$ and $S$ have maximally polarising
effects, whereas intermediate states induce flips of neighbouring
spins only at a probabilistic rate. For $S=2$ the model reduces to the
voter model, while for $S=3$ we recover the $3$-state model as
discussed above.

\begin{figure}
\centerline{
\includegraphics*[width=0.3\textwidth]{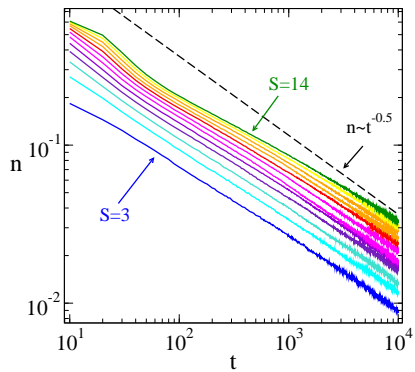}}
\caption{(Colour on-line) Coarsening of the model with multiple intermediate states. The plot shows the density of spins in intermediate states for the model at $S=3,4,...,14$ (solid lines from bottom to top). The dashed line is $n\sim t^{-0.5}$ and is shown as an optical guide only. Each curve represents data from a single run of the microscopic dynamics on a lattice of $1000^2$ sites. One unit of time corresponds to one update per spin on average. }\label{fig:multi}
\end{figure}

\begin{figure}
\centerline{
\includegraphics*[width=0.4\textwidth]{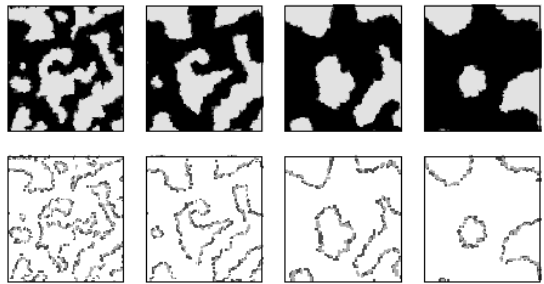}}
\caption{Snapshots of the coarsening dynamics of the model with $5$ states. Data is from simulations of a system of size $300^2$ sites, only a region of $100\times 100$ spins is shown. Snapshots are taken after $200, 400, 800$ and $1600$ sweeps respectively (left to right). In the lower row we depict only spins in the intermediate states.}\label{fig:multisnap}
\end{figure}

Results from simulations of this generalised model for several values
of $S$ are reported in Fig. \ref{fig:multi}, while some snapshots of
the dynamics for the model at $S=5$ are shown in
Fig. \ref{fig:multisnap}.  As seen in the figures, the coarsening of
the model is curvature-driven also when multiple intermediate states
are present, and the density of intermediate states decays with an
exponent close to the one already found for the $3$-state model almost
independently of the number of allowed states $S$.  In simulations
with many intermediate states ($S$ of the order of ten or more),
interfaces can be expected to have a finite size (typically ${\cal
O}(S))$. The definition of the measured quantity (density of spins in
intermediate states) can hence be affected as the number of
intermediate states is systematically increased, potentially limiting
the accuracy of our numerical results and preventing one from a
reliable measurement of the scaling exponent in simulations, even at
comparably large lattice sizes (e.g. $1000^2$). Still, results reported in Fig. \ref{fig:multi} are consistent with algebraic coarsening at an exponent of $0.45$ to $0.5$.

Other definitions of multi-state models are of course possible. For
instance one could consider a model with $S$ states (again with an
internal ordering between them) with two extremal absorbing states
($1$ and $S$), but with uniform transition probabilities among
neighboring states (e.g. a given spin in state $s$ may change to state
$s+1$ upon interaction with a spin in any state $s'>s$, with a rate
which is independent of the specific state $s'$). We have tested such
models (results not shown here) and have found algebraic domain growth
with a coarsening exponent of about $0.45$, so that we conclude that
the precise definition of the transition rates between internal states
is immaterial as far as the growth of the characteristic length scale
of the resulting domain structure is concerned.  \\

\section{Continuum description in the scaling regime}

In order to complement simulation results and to obtain analytical
insight into the dynamics of the voter model with intermediate states
we develop a description of the $3$-state model in terms of continuous
field variables.  As the microscopic 
model is neither diluted nor endowed with mobility, a standard derivation 
of continuous equations for coarse-grained degrees of freedom \cite{DFL85,MD99} is not straightforward. For the same reasons and because of the fact that the model considered here is effectively a spin-$1$ model, an exact microscopic mapping to field theoretic methods using second quantization, e.g. following the lines of \cite{C96,CT96}, is difficult as well. \\

We will therefore proceed to derive approximate stochastic evolution
equations directly from the microscopic rates, based on the assumption
that fields are Gaussian random variables. Because of the nature of
the derivation, these stochastic equations do not describe
coarse-grained degrees of freedom, but microscopic ones and differ
from those obtained by adding diffusion to mean-field like terms
\cite{MD99}.  The macroscopic description becomes clear when the slow
modes are identified, by applying power counting and neglecting higher-order terms in the momentum space.  This approach is somewhat similar
to the derivation of effective field theories in condensed matter spin
models.

We start by considering a version of the model, defined as before on a $2$-dimensional regular lattice, but now endowed with $\Omega$ spins per site, and in which each spin can take any of the three states $\{-1,0,1\}$. The dynamics at each time step then proceeds by selecting one site and one of the $\Omega$ spins at that site at random, and then executing the interaction with a randomly chosen spin at a randomly selected nearest neighbour site. The $\Omega$ copies of the system hence become cross-correlated.

In the limit $\Omega\to\infty$ deterministic equations for the evolution of average local quantities
such as the average local magnetization ($\langle s_{i} \rangle$) and
the average local density of $0$-states ($1-\langle s_{i}^{2}
\rangle$), can be easily obtained with the same rate equation approach
applied by Krapivsky in the case of pure voter model
\cite{K92,FK96}. Averages $\langle \dots\rangle$ are here understood
as averages over the $\Omega$ copies of the microscopic dynamics, i.e. $\langle s_i\rangle$ is the average magnetisation at site $i$ resulting from the $\Omega$ spins at that site. In the limit of an {\em infinite} number of particles per site we replace $\langle
s_{i} \rangle \to \overline{\phi}(x,t)$ and $1-\langle s_{i}^{2}
\rangle \to \overline{\psi}(x,t)$. Using such continuous variables is
appealing but dangerous as it may introduce ultraviolet divergencies
as well as a certain level of arbitrariness in the definition of the
continuum limit of local operators (like the lattice
laplacian). During the course of the calculations presented in this
paper we will not consider a continuum limit in space, and the
variable $x$ is assumed to have the granularity of an arbitrary
lattice spacing as in standard lattice field theories. We expect that
the spatial continuum limit could be in principle considered
appropriately defining a diffusion constant $D$.

If only a {\em finite} number $\Omega$ of spins per site is considered, then progress can be made starting from the master equation describing this ensemble of $\Omega$ interacting copies of the system. Let us label the state of spin $i$ in copy $k$ of the system by $s_i^{(k)}(t)$ at time $t$. We will denote averages over the $\Omega$ copies by $\widetilde \phi(x,t)$ in the following, i.e. we will use the substitution $\Omega^{-1}\sum_{k=1}^\Omega s_i^{(k)}(t)\to \widetilde \phi(x,t)$. Note that while overbars indicate averages in a system with an {\em infinite} number of spins per site, averages denoted by a tilde are over a {\em finite} number $\Omega$ of copies only. It is then easy to verify that a change in the state of spin $i$ in one of the $\Omega$ samples corresponds to a transition of the type, $\widetilde{\phi}(x,t) \to \widetilde{\phi}(x,t) \pm \frac{1}{\Omega}$ and $\widetilde{\psi}(x,t) \to
\widetilde{\psi}(x,t) \pm \frac{1}{\Omega}$ in the quantities describing the average over the $\Omega$ samples.  This leads to a master equation describing the evolution of the functional probability distribution ${\cal P}[\{\widetilde \phi(x,t),\widetilde\psi(x,t)\}]$. Expanding to next-to-leading order in $\Omega^{-1}$, we get the following
multivariate functional Fokker-Planck equation for the functional probability distribution $\mathcal{P}$
\begin{eqnarray}
\frac{\partial}{\partial t} \mathcal{P}(\widetilde{\phi},\widetilde{\psi};t) & = & 
- \frac{\delta}{\delta \widetilde{\phi}} \left[ \left( \frac{\widetilde{\phi} \widetilde{\psi}}{2} + \frac{1+\widetilde{\psi}}{2} \nabla^2 \widetilde{\phi} \right) \mathcal{P}(\widetilde{\phi},\widetilde{\psi};t) \right] -  \frac{\delta}{\delta \widetilde{\psi}} \left[ \left( -\frac{3 \widetilde{\psi}}{2} + \frac{1-\widetilde{\phi}^2}{2} - \frac{\widetilde{\phi}}{2} \nabla^2 \widetilde{\phi} \right) \mathcal{P}(\widetilde{\phi},\widetilde{\psi};t) \right]\nonumber\\
&&\hspace{-6em}+ \frac{1}{2 \Omega} \left( \frac{\delta^2}{\delta \widetilde{\phi}^2} +  \frac{\delta^2}{\delta \widetilde{\psi}^2}\right) \left[ \left( \frac{\widetilde{\psi}}{2} + \frac{1-\widetilde{\phi}^2}{2} - \frac{\widetilde{\phi}}{2} \nabla^2 \widetilde{\phi} \right) \mathcal{P}(\widetilde{\phi},\widetilde{\psi};t) \right] + \frac{1}{\Omega}  \frac{\delta^2}{\delta \widetilde{\psi} \delta \widetilde{\phi}} \left[ \left( \frac{1- 3\widetilde{\psi}}{2} \nabla^2 \widetilde{\phi} - \frac{3 \widetilde{\psi} \widetilde{\phi}}{2} \right) \mathcal{P}(\widetilde{\phi},\widetilde{\psi};t) \right], \label{eq:fpe}
\end{eqnarray}
where we have re-scaled time by a factor of $\Omega^{-1}$, i.e. $t\to
t/\Omega$.  $\nabla^2$ is here the lattice Laplacian, i.e. $\nabla^2\widetilde \phi(x,t)=\frac{1}{4}\sum_{y\in x}\{\widetilde\phi(y,t)-\widetilde\phi(x,t)\}$, where the sum over $y$ extends over the four nearest neighbours of $x$. Noise terms in the corresponding Langevin equations for
$\widetilde\phi$ and $\widetilde\psi$ are hence of order
$\Omega^{-1/2}$. They reflect the fact that averages in a model with only a {\em
finite} number $\Omega$ of spins per site still remain stochastic
quantities. Stochasticity only vanishes if $\Omega\to\infty$, and
deterministic mean-field equations for $\overline{\phi}$ and $\overline{\psi}$
are obtained (these would be Eqs. (\ref{3voter-eq}) below, with the
noise terms removed). Our approximation consists in describing the model with only one spin per site (i.e. the model introduced and discussed in the earlier sections) by the following two coupled Langevin equations
\begin{equation}\label{3voter-eq}
\begin{array}{cl}
\frac{\partial}{\partial t} \phi(x,t) &=  \frac{1}{2}\phi(x,t) \psi(x,t) +  \frac{1+\psi(x,t)}{2} \nabla^2 \phi(x,t) + \eta(x,t), \\
& \\
\frac{\partial}{\partial t} \psi(x,t) &= - \frac{3}{2}\psi(x,t) + \frac{1-\phi(x,t)^2}{2} - \frac{\phi(x,t)}{2}  \nabla^2 \phi(x,t) +  \xi(x,t),
\end{array}
\end{equation}
where $\eta(x,t)$ and $\xi(x,t)$ are Gaussian noise terms, with correlation given by
\begin{equation}\label{noise}
\begin{array}{cll}
\langle \eta(x,t) \eta(x',t') \rangle &= \langle \xi(x,t) \xi(x',t') \rangle &= \frac{1}{2} \left\{ \psi(x,t) + 1-\phi(x,t)^2 - \phi(x,t)  \nabla^2 \phi(x,t)\right\} \delta(x-x') \delta(t-t'), \\
&\\
\langle \eta(x,t) \xi(x',t') \rangle &=  \langle  \xi(x,t) \eta(x',t') \rangle &= \frac{1}{2} \left\{ -3 \psi(x,t) \phi(x,t) +  \left(1- 3\psi(x,t)\right) \nabla^2 \phi(x,t) \right\} \delta(x-x') \delta(t-t'). 
\end{array}
\end{equation}
We have therefore effectively disregarded higher-order terms in the expansion in $\Omega^{-1}$, even if the model actually corresponds to $\Omega =1$. Nevertheless, as we will show below, Eqs. (\ref{3voter-eq} - \ref{noise}) turn out to be a faithful
mapping of the original dynamics of the $3$-state model onto a lattice
theory with continuous fields at each grid point, capturing the essential feature of the original model.  It is here worth
stressing again that -similar to the voter model- our microscopic
dynamics do not allow for microscopic mobility of agents (e.g. hopping
or interchange of neighbouring particles), thus the diffusion-like
terms in (\ref{3voter-eq}) are non-trivial features emergent from the
microscopic model. In particular the diffusive terms in the deterministic parts of (\ref{3voter-eq}) are quadratic in the fields (e.g. $\psi\nabla^2\phi$ and $\phi\nabla^2\phi$). Similar structures have been found in the context of other spatial systems for example in \cite{McKane}.\\
A study of a version of our model including microscopic
mobility can be performed along the lines of \cite{RMF07}, here local mixing is ensured by scaling the hopping rate suitably with the system size, and an effective description in terms of continuous fields can be obtained by means of an expansion in the inverse system size. This leads to equations which are 
similar to but not identical with Eqs. (\ref{3voter-eq}) and (\ref{noise}) \cite{GDprep}. More specifically one finds noise amplitudes of the form (\ref{noise}) with the terms proportional to $\nabla^2\phi$ removed and with an additional overall prefactor $N^{-1}$ ($N=L^2$ being the system size). The deterministic terms take a form very similar to those reported in (\ref{3voter-eq}), but with decoupled diffusive terms ($ \propto \nabla^2\phi$, $\nabla^2\psi$) instead of objects of the form $\psi\nabla^2\phi$ and $\phi\nabla^2\phi$. \\ If we were to
repeat the same procedure as above for the case of the pure voter
model with two possible states (and without mobility or particle interchange), the resulting equation would be
\begin{equation}\label{eq:voter-eq0}
\frac{\partial}{\partial t} \phi(x,t) = \nabla^2 \phi(x,t) + \eta(x,t) , 
\end{equation}
where $\langle \eta(x,t) \eta(x',t') \rangle = \left\{ 1- \phi(x,t)^2
-2 \phi(x,t)  \nabla^2 \phi(x,t) \right\}\delta(x-x')
\delta(t-t')$. Apart from the diffusion term in the noise,
Eq. (\ref{eq:voter-eq0}) is identical to the field-theoretic equation
proposed by Dickman et al. for the voter model \cite{DT95}. Naive power
counting on dimensional grounds reveals that the diffusive terms involving
 Laplacians in the noise correlator are irrelevant in any dimension in the long wavelength limit. Long-time large-scale properties of a microscopic voter
model are therefore well described by the effective Langevin equation
\begin{equation}\label{voter-eq}
\frac{\partial}{\partial t} \phi(x,t) = D \nabla^2 \phi(x,t) + \sqrt{1-{\phi(x,t)}^2}\eta(x,t) , 
\end{equation}
with $\langle \eta(x,t) \eta(x',t') \rangle = \delta(x-x')
\delta(t-t')$. Note that the same equation could have been obtained also by 
adding the diffusive term to a zero-dimensional mean-field Langevin structure  \cite{DT95}. 
Despite its validity, such coarse-grained derivation misses a clear microscopic justification. \\
Similarly, on the base of dimensional analysis 
we could neglect the terms proportional to $\nabla^2\phi$ in Eq. (\ref{noise}), 
providing slightly simplified noise terms with the same absorbing states.

As we will see, numerical integration of Eqs (\ref{3voter-eq}) and
(\ref{noise}) is challenging (simple forward integration in discrete time steps renders the field-dependent variances of the noise terms negative with finite probability), therefore it is
useful to further reduce their complexity by means of some physically
meaningful approximation.\\ Provided we focus on the asymptotic
dynamics of the system, it is possible to reduce
Eqs. (\ref{3voter-eq}) and (\ref{noise}) to a single equation. We here
exploit the observation that intermediate states cannot proliferate.
Due to their immediate decay, $0$-states localize at the interfaces of
large domains of spins in the states $\pm 1$. Hence the field
$\psi(x,t)$ indicating the local presence of intermediate states is
zero if $\phi(x,t)$ is close to $-1$ or $1$, and non-zero only when $\phi(x,t)$ takes values far from the absorbing boundaries. The
approximation $\psi(x,t) \simeq 1-\phi(x,t)^2$ is thus justified and a
single field $\phi$ is expected to be sufficient to describe the
behavior of domain walls driving the asymptotic coarsening. Within
this approximation, and neglecting the modulation of the diffusion
constant induced by the term $\frac{2-\phi(x,t)^2}{2}$ in front of the
Laplacian in the first equation of (\ref{3voter-eq}) (this term takes
values near $1$ close to the interfaces and $1/2$ in the bulk), the
system is described by the following equation for the local
magnetization field $\phi$,
\begin{equation}\label{eq:3voter-eq2}
\frac{\partial}{\partial t} \phi(x,t) = \frac{1}{2}\left[\phi(x,t)- \phi(x,t)^3\right] + \nabla^2 \phi(x,t) + \sqrt{1-\phi(x,t)^2} \eta(x,t) ,
\end{equation}
where $\eta$ is the usual Gaussian noise, uncorrelated both in space
and time and of unit variance.  The first term on the right-hand-side
of Eq. (\ref{eq:3voter-eq2}), absent in the voter model, is the
typical drift term related to a double-well potential in the sense of
a Ginzburg-Landau model. It drives the system to relax towards the two
minima of the potential at $\pm 1$, coinciding with the absorbing
states of the $3$-state model.
 
In the previous section we have seen that the algebraic nature of the
decay of the interface density and the corresponding coarsening
exponent are not affected significantly when the number of
intermediate states is increased, provided that we use a model in
which interfaces remain compact (e.g. when the transition rules favor
the convergence to the absorbing states).  For the multi-state model
proposed, we expect Eq. (\ref{eq:3voter-eq2}) to give a valid
asymptotic effective description in the limit of continuous spins.  To
this end we study the limit in which the number of intermediate states
$S$ is large and in which we can approximate spin configurations by a continuum
profile. Specifically, consider $\rho = \frac{s-1}{S-1}$, where the
spin state $s$ takes values $s\in\{1,\dots,S\}$. In the limit $S\gg 1$,
$\rho$ can then be thought of as a continuous degree of freedom (at each lattice point),
taking values $\rho\in [0,1]$. In order to recover a well-defined
model it is then necessary to re-scale the spatial dimension
appropriately, viz. $x \to x / S-1$. Uniform profiles at $\rho=0$ or
$\rho=1$ are then absorbing states.  In this limit the probability
that the field $\rho_i$ at site $i$ increases (by an infinitesimal
amount) at any given time step is then linear in the value $\rho_{j}$
of one randomly chosen neighbor with which interaction occurs.\\ The
continuous representation provides a good long-time coarse-grained
approximation for a model with large $S$ in the coarsening regime,
where $L(t)$ is much larger than the interface width.  The
coarse-grained nature of this representation allows one now to derive
a phenomenological Langevin equation for the variable
$\rho(x,t)$. Considering a Fokker-Planck equation for a homogeneous
mean-field variable $\rho$ and imposing the presence of absorbing
states at $\rho=0$ and $\rho=1$, we find
\begin{equation}
\label{FP-rho}
\frac{\partial}{\partial t} P(\rho,t) = \left( 1- 2\bar{\rho} \right) \frac{\partial}{\partial \rho} \left[ \rho (1-\rho) P(\rho,t) \right] + \frac{\gamma}{2} \frac{\partial^2}{{\partial \rho}^2} \left[\rho (1-\rho) P(\rho,t) \right] , 
\end{equation}
where $\bar{\rho} = \int_{0}^{1} dy y P(y)$ is the average value 
and where $\gamma$ relates to the amount by which fields are incremented or reduced in the course of a microscopic step. The corresponding Langevin equation for the spatial dependent field $\rho(x,t)$  then reads 
\begin{equation}
\label{Langevin-rho}
\frac{\partial \rho}{\partial t}  = - ( 1- 2 \rho )\rho (1-\rho) + D \nabla^2 \rho  + \sqrt{\gamma \rho (1-\rho)} \eta , 
\end{equation}
where we have added a diffusion-like term. The noise $\eta$ is delta-correlated in time and space.
The substitution $\phi(x,t) = 1 -2\rho(x,t)$ in Eq. (\ref{Langevin-rho})
 yields a Langevin equation of the same form as Eq. (\ref{eq:3voter-eq2}).
While no claim is made that this derivation is rigorous or exact, these simple arguments lead one to expect that the entire class of voter models with $S$ intermediate states exhibits an asymptotic coarse-grained dynamics described by Eq. (\ref{eq:3voter-eq2}). Since Eq. (\ref{eq:3voter-eq2}) leads to algebraic coarsening as discussed below, this is consistent with our numerical findings for the multi-state model (see Fig. \ref{fig:multi}).

\section{Analysis of the equations}

\subsection{Field theoretic analysis and relation to the GV class}
In this section we analyze the field theoretic counterpart of this model,
that can be generally identified starting from the equation
\begin{equation}\label{eq:eqphi}
\frac{\partial}{\partial t} \phi(x,t) = a\phi(x,t)- a\phi(x,t)^3 + D \nabla^2 \phi(x,t) +  \sqrt{\gamma \left(1-\phi(x,t)^2\right)} \eta(x,t) ,
\end{equation}
i.e. a special case of the Langevin equation of the Generalized Voter class proposed by Al Hammal et al. in
Ref. \cite{ACDM05} (one has $b=0$, $a>0$ in their notation).
The fact that linear and cubic terms have the same coupling constant is necessary to enforce $\mathbb{Z}_{2}$-symmetry and has striking consequences for the corresponding field theory.\\
Using the response functional technique \cite{C96}, we obtain the 
following effective action
\begin{equation}\label{action}
\mathcal{S}[\phi,\hat{\phi}] = \int d^{d}x dt \left[ {\hat{\phi}\left( \partial_{t} -a -D\nabla^2 \right) \phi + a\hat{\phi} \phi^3 - \gamma \hat{\phi}^2 +  \gamma \hat{\phi}^2 \phi^2}\right]~.
\end{equation}  
Naive power counting reveals the canonical dimensions of the fields,
$[\phi] = 1$ and $[\hat{\phi}]=k^d$ and that of coupling constants,
$[a]=k^2$ and $[\gamma]=k^{2-d}$.  The noise vertices become relevant
for $d \leq 2$, whereas for $d>2$ the process is not affected by
multiplicative noise.  In fact, the critical theory ($a=0$, i.e. voter
model) can be renormalized exactly to all orders, leading to a
complete characterization of the ordering process in any dimension,
above and below $d_{c}=2$ \cite{P86,L94}.\\ For any $a>0$, the system
is in the broken symmetry phase with spontaneous magnetization and
minima in $\pm 1$. In analogy with model A below the critical
temperature, the dynamical field theory should be governed by the
strong coupling fixed point $a \to \infty$ \cite{B90,B94}.  This can
be verified by developing momentum shell renormalization for the
coupling $a$. Under rescaling of lengths and times, $x \to b x$, $t
\to b^{z} t$, one considers the transformations $D \to b^{z-2} D$, $a
\to b^{z} a$ and $\gamma \to b^{z-d} \gamma$. Imposing scale
invariance for the system's fluctuations, as required by coarsening
regime, we get $z=2$ and the tree-level flow equations $\frac{d
a}{d\ell} = 2 a $, and $\frac{d \gamma}{d\ell} = (2-d) \gamma$, where
$b \equiv e^{\ell}$.  These simple renormalisation group (RG)
equations hence predict a strong coupling fixed point for $a$, playing
the same role as the zero-temperature fixed point does in the Ising
model.  This means that in the region $a>0$ we do not expect
corrections to the mean-field dynamical exponent $z=2$, so that this
exponent, characterising the growth of the correlation length during
coarsening, is expected to be accurate in any dimension. This is also
true for $a=0$ since the propagator does not acquire corrections.  The
density of interfaces on the other hand follows the general scaling
$n(t) \sim 1/L(t) \sim t^{-1/z}$ in any dimension only for $a>0$,
whereas it is modified to a logarithmic behavior for the critical
(voter) theory ($a=0$).\\ We have here disregarded corrections to the
tree level RG equations and have neglected higher-order diagrams. The
absence of loop corrections for $a>0$ can be seen to be justified by
performing a mapping of this model on branching annihilation random
walks (BARW) by rapidity inversion \cite{CT96,CCDDM05}.  Our model
here corresponds to unphysical BARW with negative reaction rates.
However, following \cite{CT96}, loop corrections shift the critical
value at $a=0$ to some $a_{c} <0$, whereas our discussion concerns the
absorbing phase at $a>0$.  Note that region $a<0$ does not correspond
to any microscopic model obtained adding intermediate states to the
voter model, but presents a non-trivial competition between the two
absorbing states at $\pm 1$ and the minimum of the potential at zero
\cite{ACDM05}.  For this case, a complete understanding of the RG
diagram is still lacking \cite{CT96,CCDDM05,DCM05}.

The fact that noise flows to zero for $d>2$ and to the strong coupling fixed point for $d<2$ indicates that roughening phenomena may be present below the upper critical dimension $d_{c} =2$, but that they are absent in $d>2$. In dimensions higher than two the system therefore orders like a zero-temperature Ising model. The behavior at $d=2$ can be understood by mapping the field theory on the dynamics of an interface using standard methods \cite{BS95}. Consider the Hamiltonian formulation for the equilibrium problem in absence of multiplicative noise $\mathcal{H} = \int \left[\frac{1}{2}{\left(\nabla \phi \right)}^2 + V(\phi)\right] d^{d}x$, where $V(\phi) = \frac{a}{4} (1-\phi^2)^2$, and define the interface position is $z = h(y,t)$, with $y$ being the $d-1$ dimensional space coordinate specified by defining a preferred direction $z$ for the interface movement (i.e. $\vec{x} = (\vec{y},\vec{z})$). Neglecting the multiplicative noise, it is a zero temperature Ising model, so the interfaces will be mainly driven by diffusive forces \cite{J84}. On the other hand, the multiplicative noise term behaves like additive noise at the interface (since $\phi = 0$ near the interface), resulting in the following Edwards-Wilkinson (EW) type of equation,
\begin{equation} 
\frac{\partial h(y,t)}{\partial t} = \tilde{D} \nabla_{\perp}^2 h(y,t) + \eta(y,t) 
\end{equation}
with $\langle \eta(y,t) \eta(y',t') \rangle = \gamma \delta(t-t') \delta^{d-1}(y-y')$.
Therefore the interface roughening in $d \leq 2$ is described by the EW universality class in $d_{int}=d-1$ dimensions. In $d_{int}=1$, EW universality class predicts roughening exponents $\alpha = 1 - d_{int}/2 = 1/2$ and $\beta = \alpha/z = 1/4$, consistent with the behavior of the interface width $W$ measured for our model in simulations (see Fig. \ref{fig:roughness}).

Interestingly, we could also apply an infinitesimal magnetic field in order to break the $\mathbb{Z}_{2}$-symmetry, then expanding the fields around one of the absorbing states (e.g. $\phi= \phi_{0} - \phi'$ with $\phi_{0}=1$)\cite{P98}.
In this case, the resulting action $\mathcal{S'}$ for the perturbation field $\phi'$ reads
\begin{equation}\label{pert-action}
\mathcal{S'}[\phi',\hat{\phi'}] = \int d^{d}x dt \left[ {\hat{\phi'}\left( 
\partial_{t} + 2a -D\nabla^2 \right) \phi' - 3a\hat{\phi'} {\phi'}^2  + 2 \gamma \hat{\phi'}^2 \phi'} - \gamma \hat{\phi'}^2 {\phi'}^2 \right]~,
\end{equation}  
in which we have neglected higher order terms in $\phi'$.\\
Eq. (\ref{pert-action}) is similar to the action of directed percolation (DP) \cite{H00}, therefore it is convenient to put it in the standard DP form \cite{THVL05} by rescaling the fields. Let us define $\hat{\zeta} = \sqrt{3a/2\gamma} \hat{\phi'}$ and $\zeta = \sqrt{2\gamma/3a} \phi'$, after a bit of algebra we get
\begin{equation}\label{pert-action2}
\mathcal{S'}[\zeta,\hat{\zeta}] = \int d^{d}x dt \left[ \hat{\zeta}\left( 
\partial_{t} + 2a -D\nabla^2 \right) \zeta - u \left(\zeta - \hat{\zeta}\right) \hat{\zeta} \zeta  - \gamma \hat{\zeta}^2 {\zeta}^2 \right]~,
\end{equation}     
where $u = \sqrt{6 a \gamma}$. Like in DP, the pure noise coupling $\gamma$ can be neglected with respect to $u$, since $[u]=k^{2-d/2}$ and $[\gamma/u]=k^{-d/2}$, thus the effective action becomes 
\begin{equation}\label{pert-action3}
\mathcal{S}_{eff}[\zeta,\hat{\zeta}] = \int d^{d}x dt \left[ \hat{\zeta}\left(\partial_{t} + 2a -D\nabla^2 \right) \zeta - u \left(\zeta - \hat{\zeta}\right) \hat{\zeta} \zeta \right]~.
\end{equation} 
It is easy to check that one loop correction to the propagator shift the bare critical point ($a=0$) outside the region of interest (i.e. towards negative values $a_{R} <0$). 
In other words, when we apply a small magnetic field to our system, the corresponding field theory should be that of directed percolation in the absorbing or active phase depending on the sign of $a$. In terms of interface dynamics, the presence of a small magnetic field breaking the $\mathbb{Z}_{2}$-symmetry should bring the system in the Kardar-Parisi-Zhang universality class \cite{BS95}. It would here probably be desirable to have a microscopic model with $S\geq 3$ states available to investigate such cases with broken $\mathbb{Z}_2$-symmetry further.

\subsection{Numerical integration}

To verify the validity of the equations proposed in the previous
sections, and to compare their behavior with the results of
microscopic simulations, we have integrated them numerically,
considering both deterministic cases and versions of the equations
which are subject to noise.  It is well known that multiplicative
noise prevents integration by naive Euler-forward discretisation, but
various methods have been recently proposed to overcome this problem
(see e.g. \cite{DCM05,M04} and references therein).  Unfortunately, a
generalization of these step-split methods \cite{DCM05,M04} to systems
of coupled stochastic differential equations with cross-correlated
noises is still awaited, therefore in the case of
Eqs. (\ref{3voter-eq}) we limit our study to the deterministic variant
of the equations.  Slight dependencies on e.g. system size and
time-discretisation in the numerical integration of these equations
are here hard to eliminate due to limitations in computing time (we
have used a standard Euler-forward scheme at time stepping of
$0.01$). Nevertheless, the results for the density of interfaces
$n\sim1-\langle \phi^2 \rangle$, reported in Fig. \ref{fig:micro}, are
in good agreement with measurements from microscopic simulations, indicating a
scaling exponent of about $0.45$ to $0.5$. They are also consistent with
the snapshots generated from the microscopic dynamics (see
Fig. \ref{fig:droplet}) in which a curvature-driven coarsening is
observed, even if domain boundaries are not completely smooth due to
the interfacial roughening discussed above.  This provides further
evidence that noise cannot be expected to play a relevant role in the
asymptotic coarsening dynamics of the voter model with intermediate
states.\\ In the case of Eq. (\ref{eq:3voter-eq2}) the step-split
method can be applied and both stochastic and deterministic versions
of this reduced equation can hence be studied numerically. Split-step
methods for the stochastic case here rely on the fact that the
Fokker-Planck equation for stochastic processes of the form
$\dot\phi=\sqrt{\phi}\eta$ can be solved exactly, and that the
resulting field distributions can subsequently be sampled. We here
employ the scheme of \cite{M04}, using the replacement
$\sqrt{1-\phi^2}\to\Theta(\phi)\sqrt{1-\phi}+\Theta(-\phi)\sqrt{1+\phi}$
as described in \cite{DCM05}. The deterministic remainder of the
equation is then integrated using a simple Euler-forward
scheme. Typical time steps used are here of the order of $0.25$ \cite{M04,DCM05}. Results are shown in Fig. \ref{fig:micro} for different values
of the noise amplitude $\gamma$. Although the quantitative details of
these numerical findings show dependencies on parameters such as
system size or time-stepping, results are consistent with an exponent
of $0.45$ to $0.5$ for the decay of the density of interfaces, and
hence in good agreement with the above theoretical considerations and
direct measurements from simulations of the microscopic dynamics of
the $3$-state model. At variance with the analogous equation
describing the standard voter model without intermediate states
\cite{DCM05}, and in agreement with the field theoretic prediction,
the scaling behaviour of the coarsening process is not significantly
affected by noise amplitude $\gamma$ in Eq. (\ref{eq:eqphi}).  In
conclusion, numerical data obtained from integration of
Eqs. (\ref{3voter-eq}) and (\ref{eq:3voter-eq2}) corroborate the
theoretical analysis and demonstrate the validity of the continuous
approach to describe long time coarsening behavior of the microscopic
voter model with intermediate states.  The fact that both
Eqs. (\ref{3voter-eq}) and Eqs. (\ref{eq:3voter-eq2},\ref{eq:eqphi}) give an exponent
$\delta \approx 0.45$ and that this value does not change much with
the amplitude of the noise, leads one to suggest that the discrepancy
with the theoretically expected value $0.5$ of Model A dynamics is
probably due to finite size effects in the simulations, or to other
numerical inaccuracies.

\section{Conclusions}
The voter model is one of the most studied microscopic models of
interacting particle systems \cite{L85}, and a paradigm for
phase-ordering processes driven by interfacial noise. Recent
developments in the study of non-equilibrium statistical physics have
put forward a common paradigm to study coarsening with and without
surface tension, which has led to the identification of the {\em
generalized voter} (GV) universality class \cite{ACDM05}. Voter models
with modified spin-flip probability (i.e. breaking the linearity with
the local field) have been proved to belong to this general class
\cite{DK84,OMS93,DG99,DCCH01}. In this paper we have shown that voter
models with intermediate states fall into this class as well, and have
studied their scaling behavior in detail. In particular, we observe
algebraic domain ordering and the density of interfaces is found to
decay with an exponent of $0.45$ to $0.5$. \\ We have derived a set of
two coupled stochastic differential equations for appropriately
defined fields, and have demonstrated that these equations faithfully
describe the long-time large-scale behavior of the $3$-state voter
model. Within a physically motivated approximation the two equations
can be seen to reduce to one equation for a single field, and the
latter equation in turn takes a form of a special case of the GV
class. Moreover, the same algebraic coarsening is found in models with
more than one intermediate state, and similar equations appear to hold
for such cases with $S>3$.\\ We have used a field theoretical approach
to understand the universal properties of the model and have
demonstrated that numerical results (obtained from both microscopic
simulations and numerical integration of the effective equations) are
consistent with this theoretical analysis.  Interestingly we find
that, due to the appearance of an effective surface tension induced by
the presence of intermediate states, the domains coarsen like in the
low temperature phase of Model A dynamics. In $d \leq 2$, the presence
of interfacial noise induce interfacial roughening in agreement with
usual Edwards-Wilkinson scaling. RG equations predict smooth
interfaces in higher dimension. \\ It is hoped that our theoretical
analysis will help to shed light on the phenomenology of this class of
generalised voter models. These models have come to the attention of
statistical physicists only recently, mostly due to their applications
in the modelling of social systems. With a modest amount of
imagination however, they can be expected to be applicable also in the study
of kinetic reactions with intermediate unstable compounds and in other
physical, biological or chemical non-equilibrium systems.

\vspace{1em}
{\em Acknowledgements:} The authors would like to thank A. J. Bray, C. Castellano, M. Marsili and E. Moro for helpful discussions. 
This work is supported by an RCUK Fellowship (RCUK reference EP/E500048/1).

\end{document}